\documentstyle[12pt,epsf]{article}

\def\spose#1{\hbox to 0pt{#1\hss}}
\def\lsim{\mathrel{\spose{\lower 3pt\hbox{$\mathchar"218$}}
 \raise 2.0pt\hbox{$\mathchar"13C$}}}
\def\gsim{\mathrel{\spose{\lower 3pt\hbox{$\mathchar"218$}}
 \raise 2.0pt\hbox{$\mathchar"13E$}}}

\begin{document}

\begin{titlepage}

\begin{flushright}
CERN-TH/97-148\\
hep-ph/9707217
\end{flushright}

\vspace{1.5cm}
\begin{center}
\Large\bf
Theory of Beauty Lifetimes
\end{center}

\vspace{0.5cm}
\begin{center}
M. Neubert\\[0.1cm]
{\sl Theory Division, CERN, CH-1211 Geneva 23, Switzerland}
\end{center}

\vspace{1cm}
\begin{abstract}
\vspace{0.2cm}\noindent
A critical review of the theoretical understanding of the lifetimes of
beauty hadrons is given. A model-independent analysis using the
heavy-quark expansion allows for a description of the lifetime ratios
$\tau(B^-)/\tau(B^0)$ and $\tau(\Lambda_b)/\tau(B^0)$ in a small region
of parameter space. It is demonstrated that the lifetime ratio
$\tau(B^-)/\tau(B^0)$ cannot be used to extract the decay constant of
the $B$ meson. Implications for the semileptonic branching ratio and
charm yield in $B$ decays are pointed out.
\end{abstract}

\vspace{1.5cm}
\begin{center}
Invited talk presented at the\\
Second International Conference on $B$ Physics and CP Violation\\
Honolulu, Hawaii, 24--27 March 1997
\end{center}

\vspace{2cm}
\noindent
CERN-TH/97-148\\
July 1997

\end{titlepage}

\section{Introduction}

The theoretical description of inclusive decays of hadrons containing a
heavy quark exploits two observations~\cite{Chay}--\cite{shape}:
bound-state effects related to the initial state can be calculated
using the heavy-quark expansion, and the fact that the final state
consists of a sum over many hadronic channels eliminates the
sensitivity to the properties of individual final-state hadrons. The
second feature rests on the hypothesis of quark-hadron duality, i.e.\
the assumption that decay rates are calculable in QCD after a
``smearing'' procedure has been applied~\cite{PQW}. In semileptonic
decays the integration over the lepton spectrum provides a smearing
over the invariant hadronic mass of the final state (so-called global
duality), whereas for nonleptonic decays, where the total hadronic mass
is fixed, the summation over many hadronic final states provides an
averaging (so-called local duality). The calculation of the lifetimes
of heavy hadrons relies on the stronger assumption of local duality.

At present, quark-hadron duality cannot be derived from first
principles, although some interesting attempts to address the problem
have been discussed recently~\cite{Shifm,Boyd}. The validity of global
duality (at energies even lower than those relevant in $B$ decays) has
been tested experimentally using high-precision data on hadronic $\tau$
decays~\cite{Maria}. The study of beauty lifetime ratios provides a
test of local duality at energies of order a few GeV. Unlike the
semileptonic branching ratio of $B$ mesons, lifetime ratios are
insensitive to possible new physics contributions and thus probe
directly low-energy strong interactions. The common folklore is that
the current experimental value $\tau(\Lambda_b)/\tau(B)\approx 0.8$
cannot be accommodated by theory, indicating sizeable duality
violations. If this were to be true, we could not trust the predictions
of the heavy-quark expansion for other inclusive quantities, such as
the semileptonic branching ratio and charm yield in $B$ decays, which
are considered sensitive probes of new physics. A calibration of the
theoretical uncertainties in the predictions for these quantities is
therefore most important.

Here we shall reconsider the theoretical description of beauty lifetime
ratios taking a conservative point of view, which avoids making
uncontrolled approximations~\cite{MNChris}. We concentrate on the
ratios $\tau(B^-)/\tau(B^0)$ and $\tau(\Lambda_b)/\tau(B^0)$, for which
precise experimental data exist. Our model-inde\-pen\-dent approach is
less ambitious than previous analyses in that it gives up some
predictive power. Yet, we find it important to question commonly made
assumptions before concluding a failure of the heavy-quark expansion.
We find that at present it is still possible to reproduce the
experimental data on beauty lifetimes in the framework of the
heavy-quark expansion without invoking duality violations. Only a
better control of theoretical uncertainties could help to clarify the
situation.

\section{Heavy-Quark Expansion}

\begin{figure}[t]
\epsfxsize=10cm
\centerline{\epsffile{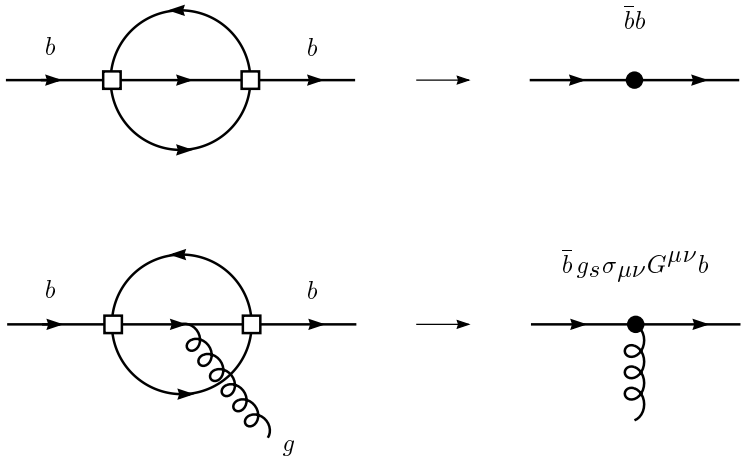}}
\caption{\label{fig:Toper}
Leading contributions to the transition operator ${\bf T}$ (left), and
the corresponding operators in the OPE (right). The open squares
represent a four-fermion interaction of the effective weak Lagrangian.}
\end{figure}

The inclusive decay width of a hadron $H_b$ containing a $b$ quark can
be written in the form
\begin{equation}\label{ImT}
   \Gamma(H_b\to X) = \frac{1}{M_{H_b}}\,\mbox{Im}\,
   \langle H_b|\,{\bf T}\,|H_b\rangle \,,
\end{equation}
where the transition operator ${\bf T}$ is given by
\begin{equation}
   {\bf T} = i\!\int{\rm d}^4x\,T\{\,
   {\cal L}_{\rm eff}(x),{\cal L}_{\rm eff}(0)\,\} \,.
\end{equation}
Here ${\cal L}_{\rm eff}$ is the effective weak Lagrangian renormalized
at the scale $m_b$~\cite{AltM,Gail}. The leading contributions to the
transition operator are given by the two-loop diagrams shown on the
left-hand side in Fig.~\ref{fig:Toper}. Because of the large mass of
the $b$ quark, the momenta flowing through the internal propagator
lines are large. It is thus possible to construct an operator product
expansion (OPE) for the transition operator, in which it is represented
as a series of local operators containing the $b$-quark fields. The
operator with the lowest dimension is $\bar b b$. It arises by
contracting the internal lines of the first diagram. The only
gauge-invariant operator with dimension 4 is $\bar b\,i\rlap{\,/}D\,b$;
however, the equations of motion imply that this operator can be
replaced by $m_b\bar b b$. The first operator that is different from
$\bar b b$ has dimension 5 and contains the gluon field. It arises from
diagrams in which a gluon is emitted from one of the internal lines,
such as the second diagram shown in Fig.~\ref{fig:Toper}. From
dimension 6 on, a large number of operators appears. For dimensional
reasons, the matrix elements of higher-dimensional operators are
suppressed by inverse powers of the $b$-quark mass. Thus, the total
inclusive decay rate (i.e.\ the inverse lifetime) of a hadron $H_b$ can
be written as~\cite{Bigi,MaWe}
\begin{equation}\label{gener}
   \Gamma(H_b) = \frac{G_F^2 m_b^5 |V_{cb}|^2}{192\pi^3}\,
   \left\{ c_3\,\langle\bar b b\rangle
   + c_5\,\frac{\langle\bar b\,g_s\sigma_{\mu\nu} G^{\mu\nu} b\rangle}
               {m_b^2}
   + \sum_n c_6^{(n)}\,\frac{\langle O_6^{(n)}\rangle}{m_b^3}
   + \dots \right\} \,,
\end{equation}
where the prefactor arises from the loop integrations, $c_i$ are
calculable coefficient functions, and $\langle O\rangle$ are the
(normalized) forward matrix elements between $H_b$ states. These matrix
elements can be systematically expanded in powers of $1/m_b$ using the
heavy-quark effective theory (HQET)~\cite{review}. The result
is~\cite{Bigi,MaWe}
\begin{eqnarray}
   \langle\bar b b\rangle &=& 1
    - \frac{\mu_\pi^2(H_b)-\mu_G^2(H_b)}{2 m_b^2} + O(1/m_b^3) \,,
    \nonumber\\
   \langle\bar b\,g_s\sigma_{\mu\nu} G^{\mu\nu} b\rangle
   &=& 2\mu_G^2(H_b) + O(1/m_b) \,,
\end{eqnarray}

\vspace{-0.1cm}
\noindent
where $\mu_\pi^2(H_b)$ and $\mu_G^2(H_b)$ are the matrix elements of
the heavy-quark kinetic energy and chromomagnetic interaction inside
the hadron $H_b$, respectively~\cite{FaNe}. Inserting these relations
into (\ref{gener}) and taking the ratio of two lifetimes
yields~\cite{MNChris}
\begin{eqnarray}\label{taucrude}
   \frac{\tau(B^-)}{\tau(B^0)} &=& 1 + O(1/m_b^3) \,, \nonumber\\
   \frac{\tau(\Lambda_b)}{\tau(B^0)} &=& 1
    + \frac{\mu_\pi^2(\Lambda_b)-\mu_\pi^2(B)}{2 m_b^2}
    + c_G\,\frac{\mu_G^2(\Lambda_b)-\mu_G^2(B)}{m_b^2}
    + O(1/m_b^3) \nonumber\\
   &=& (0.98\pm 0.01) + O(1/m_b^3) \,,
\end{eqnarray}
where $c_G=-\frac 12-2c_5/c_3\approx 1.2$, and isospin symmetry is
assumed for the matrix elements of $B^-$ and $B^0$ mesons. The hadronic
parameters appearing at order $1/m_b^2$ in the ratio
$\tau(\Lambda_b)/\tau(B^0)$ can all be determined from spectroscopy.
The matrix elements of the chromomagnetic interaction are given by
\begin{equation}
   \mu_G^2(B) = \frac 34 (M_{B^*}^2 - M_B^2)
   \approx 0.36\,\mbox{GeV}^2 \,, \qquad
   \mu_G^2(\Lambda_b) = 0 \,,
\end{equation}
where the second relation follows from the fact that in the
ground-state $\Lambda_b$ baryon the light constituents have total spin
zero. The individual matrix elements of the kinetic energy are not
physical quantities because of renormalon ambiguities~\cite{renorm};
however, their difference obeys the relation
\begin{equation}
   \mu_\pi^2(\Lambda_b) - \mu_\pi^2(B) = - \frac{M_B M_D}{2}
   \left( \frac{M_{\Lambda_b}-M_{\Lambda_c}}{M_B-M_D}
   - \frac 34\,\frac{M_{B^*}-M_{D^*}}{M_B-M_D} - \frac 14 \right) \,,
\end{equation}
which implies $\mu_\pi^2(\Lambda_b)=\mu_\pi^2(B)$ up to small
corrections of order $0.01\,\mbox{GeV}^2$.

\begin{figure}[t]
\epsfxsize=10cm
\centerline{\epsffile{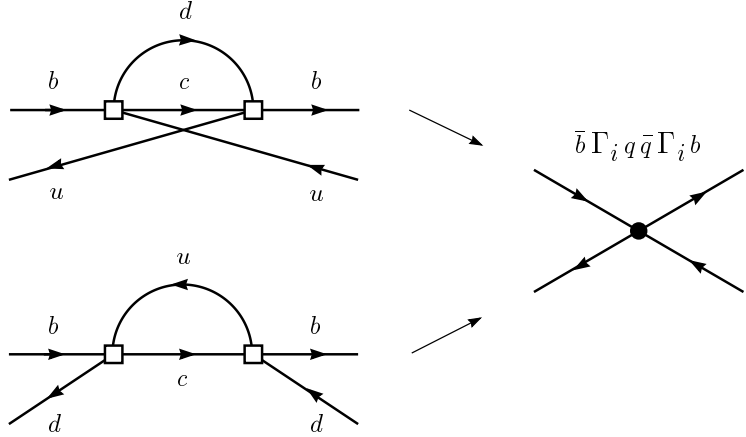}}
\caption{\label{fig:Tspec}
Nonspectator contributions to the transition operator ${\bf T}$ (left),
and the corresponding operators in the OPE (right). $\Gamma_i$ denote
combinations of Dirac and colour matrices.}
\end{figure}

The model-independent theoretical predictions in (\ref{taucrude}) may
be compared with the average experimental values for the lifetime
ratios, which are~\cite{LEPB}
\begin{equation}\label{taudata}
   \frac{\tau(B^-)}{\tau(B^0)} = 1.06\pm 0.04 \,, \qquad
   \frac{\tau(\Lambda_b)}{\tau(B^0)} = 0.79\pm 0.05 \,.
\end{equation}
Whereas the first ratio is in good agreement with the theoretical
expectation, the low value of the lifetime of the $\Lambda_b$ baryon
constitutes a puzzle. To understand the fine structure of the lifetime
ratios requires to go further in the $1/m_b$ expansion. At order
$1/m_b^3$ in the expansion of nonleptonic decay rates there appear
four-quark operators whose matrix elements explicitly depend on the
flavour of the spectator quark(s) in the hadron $H_b$, and hence
differentiate between hadrons with different light constituents. These
nonspectator effects receive a phase-space enhancement factor of
$O(16\pi^2)$ with respect to the leading terms in the OPE, which makes
them more important than the corrections of order $1/m_b^2$. This can
be seen from Fig.~\ref{fig:Tspec}, which shows that the corresponding
contributions to the transition operator arise from one-loop rather
than two-loop diagrams. In total, a set of four four-quark operators is
induced by nonspectator effects. They are
\begin{eqnarray}
   O_1 &=& \bar b\gamma_\mu (1-\gamma_5)q\,
    \bar q\gamma^\mu (1-\gamma_5)b \,, \nonumber\\
   O_2 &=& \bar b(1-\gamma_5)q\,\bar q(1+\gamma_5)b \,,
    \nonumber\\
   T_1 &=& \bar b\gamma_\mu (1-\gamma_5) t_a q\,
    \bar q\gamma^\mu (1-\gamma_5) t_a b \,, \nonumber\\
   T_2 &=& \bar b(1-\gamma_5) t_a q\,
    \bar q(1+\gamma_5) t_a b \,,
\end{eqnarray}
where $q$ is a light quark, and $t_a$ are the generators of colour
SU(3). Since hadronic matrix elements of four-quark operators are
notoriously difficult to calculate, one needs to parametrize them. For
the $B$-meson matrix elements renormalized at the scale $m_b$, we
define~\cite{MNChris}
\begin{equation}\label{Biepsi}
   \langle B|\,O_i\,|B\rangle_{\mu=m_b} \equiv B_i\,f_B^2 M_B^2
   \,, \qquad
   \langle B|\,T_i\,|B\rangle_{\mu=m_b} \equiv \varepsilon_i\,f_B^2
   M_B^2 \,,
\end{equation}
where $f_B$ is the decay constant of the $B$ meson. The values of the
dimensionless hadronic parameters $B_i$ and $\varepsilon_i$ are
currently not known. Ultimately, they may be calculated using some
field-theoretic approach such as lattice gauge theory. Some (but not
all) of these parameters may also be extracted from a precise
measurement of the lepton spectrum in the endpoint region of
semileptonic $B$ decays~\cite{BiUr}. We note that in the vacuum
saturation (or factorization) approximation~\cite{SVZ} $B_i=1$ and
$\varepsilon_i=0$ at some scale $\mu$, where the approximation is
assumed to hold. More generally, the large-$N_c$ counting rules of QCD
imply
\begin{equation}
   B_i = O(1) \,,\qquad \varepsilon_i=O(1/N_c) \,.
\label{largeNc}
\end{equation}

The theoretical expressions can also be written in terms of parameters
$B_i(\mu)$ and $\varepsilon_i(\mu)$ renormalized at a scale $\mu\ne
m_b$, using evolution equations for the hadronic matrix elements in the
HQET. Such a rewriting does not induce any uncertainty, as it is simply
a reparametrization of the results. Still, it may be appropriate to use
a lower renormalization point if model calculations of the parameters
$B_i$ and $\varepsilon_i$ are used. To give an example, we relate the
parameters in (\ref{Biepsi}) to ones renormalized at a low hadronic
scale $\mu_{\rm had}$ chosen such that $\alpha_s(\mu_{\rm had})=0.5$.
In defining these parameters, we also renormalize the decay constant
$f_B$ at the scale $\mu_{\rm had}$. Then the relations
are~\cite{MNChris}
\begin{eqnarray}
   B_i &\approx& \phantom{-}1.01 B_i(\mu_{\rm had})
    - 0.24\varepsilon_i(\mu_{\rm had}) \,, \nonumber\\
   \varepsilon_i &\approx& -0.05 B_i(\mu_{\rm had})
    + 0.73\varepsilon_i(\mu_{\rm had}) \,.
\label{evol}
\end{eqnarray}

Next consider the matrix elements of the four-quark operators between
$\Lambda_b$ baryons. Heavy-quark spin symmetry implies the relations
$\langle O_2\rangle=-\frac 12\langle O_1\rangle$ and $\langle
T_2\rangle=-\frac 12\langle T_1\rangle$~\cite{MNChris}, which reflect
the fact that the light degrees of freedom in a $\Lambda_b$ baryon are
in a state with total spin zero. This leaves two independent matrix
elements of the operators $O_1$ and $T_1$. It is convenient to
introduce the operator $\widetilde O_1=\bar b^i\gamma_\mu(1-\gamma_5)
q^j\,\bar q^j\gamma^\mu(1-\gamma_5)b^i$ instead of $T_1$ ($i,j$ are
colour indices) by making use of the Fierz identity $T_1=-\frac 16
O_1+\frac 12\widetilde O_1$. The analogue of the vacuum saturation
approximation in the case of baryons is the valence-quark assumption,
in which the colour of the quark fields in the operators is identified
with the colour of the quarks inside the baryon. Since the colour wave
function of a baryon is totally antisymmetric, the matrix elements of
$O_1$ and $\widetilde O_1$ differ in this approximation only by a sign.
Hence, it is natural to define
\begin{equation}\label{Btildef}
   \langle\Lambda_b|\,\widetilde O_1\,|\Lambda_b\rangle_{\mu=m_b}
   \equiv - \widetilde B\,\langle\Lambda_b|\,O_1\,
   |\Lambda_b\rangle_{\mu=m_b} \,,
\end{equation}
with $\widetilde B=1$ in the valence-quark approximation. For the
baryon matrix element of $O_1$ itself, we adopt a parametrization
guided by the quark model and write
\begin{equation}\label{rdef}
   \frac{1}{2 M_{\Lambda_b}}\,\langle\Lambda_b|\,O_1\,
   |\Lambda_b\rangle_{\mu=m_b}
   \equiv - \frac{f_B^2 M_B}{48}\,r \,.
\end{equation}
In the quark model $r$ is the ratio of the squares of the wave
functions determining the probability to find a light quark at the
location of the $b$ quark inside the $\Lambda_b$ baryon and the $B$
meson~\cite{Bili,ShiV}. The parameter $r$ is by far the most uncertain
one entering the theoretical formulae for beauty lifetimes. Theoretical
estimates range from $r\approx 0.1$--0.5~\cite{Gube,Cola} to $r\approx
1$--2~\cite{Rosn}. In the quark model, $r$ can be related to the ratio
of the spin splittings between $\Sigma_b$ and $\Sigma_b^*$ baryons and
$B$ and $B^*$ mesons, leading to~\cite{Rosn}
\begin{equation}
   r \approx \frac 43\,\frac{M_{\Sigma_b^*}^2 - M_{\Sigma_b}^2}
                      {M_{B^*}^2 - M_B^2} \,.
\end{equation}
Using the (still preliminary) result $M_{\Sigma_b^*}-M_{\Sigma_b} =
(56\pm 16)\,$MeV reported by the DELPHI Collaboration~\cite{DELPHI},
one then obtains $r\approx 1.8\pm 0.5$.

\section{Model-Independent Analysis of Lifetime Ratios}

Including the nonspectator contributions of order $1/m_b^3$ allows us
to refine the results in (\ref{taucrude}) at the prize of introducing
several unknown hadronic parameters. We obtain~\cite{MNChris}
\begin{eqnarray}
   \frac{\tau(B^-)}{\tau(B^0)} &=& 1 + \xi \Big[ k_1 B_1 + k_2 B_2
    + k_3 \varepsilon_1 + k_4 \varepsilon_2 \Big] \,, \nonumber\\
   \frac{\tau(\Lambda_b)}{\tau(B^0)} &=& 0.98 + \xi \Big[ p_1 B_1
    + p_2 B_2 + p_3 \varepsilon_1 + p_4 \varepsilon_2
    + (p_5 + p_6 \widetilde B) r \Big] \,,
\label{master}
\end{eqnarray}
where $\xi=(f_B/200\,\mbox{MeV})^2$. The coefficients $k_i$ and $p_i$
depend on the quark-mass ratio $m_c/m_b$ and the values of the Wilson
coefficients $c_\pm(m_b)$ appearing in the effective weak Lagrangian.
They also depend on the perturbative scheme adopted in the calculation.
So far, we have discussed the scheme in which the standard QCD
evolution is used to obtain the effective weak Lagrangian at the scale
$m_b$, where the matrix elements in the heavy-quark expansion are
evaluated. To investigate the perturbative uncertainties, one may also
renormalize the weak Lagrangian at a scale $\mu=\kappa m_b$ with
$\kappa=O(1)$, then perform the heavy-quark expansion, and finally
evolve the hadronic matrix elements from $\kappa m_b$ to $m_b$ using
evolution equations in the HQET. If it was not for the truncation of
perturbation theory, the two procedures would yield to the same result.
The dependence of $k_i$ and $p_i$ on the choice of $\kappa$ is
therefore a measure of the truncation error. In Table~\ref{tab:1}, we
give the values of these coefficients for the three choices
$\kappa=1/2$, 1 and 2. Because of an accidental cancellation between
the contribution from different operators in the effective weak
Lagrangian, the coefficients $k_1$ and $p_5$ have a significant
dependence on the choice of the ``matching'' scale, whereas the other
coefficients are either negligibly small or stable with respect to
variations of $\kappa$.

\begin{table}[htb]
\caption{\label{tab:1}
Coefficients $k_i$ and $p_i$ appearing in the predictions for
the lifetime ratios. We use $c_+(m_b)=0.86$, $c_-(m_b)=1.35$, and
$m_c/m_b=0.29$}
\vspace{0.5cm}
\centerline{\begin{tabular}{|c|cccccc|}\hline\hline
\rule[-0.15cm]{0cm}{0.5cm} $\kappa$ \rule[-0.15cm]{0cm}{0.5cm}
 & $k_1$ & $k_2$ & $k_3$ & $k_4$ & & \\
\hline
1/2 & $+0.046$ & 0.003 & $-0.737$ & 0.201 & & \\
1   & $+0.021$ & 0.004 & $-0.699$ & 0.195 & & \\
2   & $-0.007$ & 0.007 & $-0.666$ & 0.189 & & \\
\hline
\rule[-0.15cm]{0cm}{0.5cm}
 & $p_1$ & $p_2$ & $p_3$ & $p_4$ & $p_5$ & $p_6$ \\
\hline
1/2 & $-0.003$ & 0.003 & $-0.178$ & 0.201 & $-0.017$ & $-0.023$ \\
1   & $-0.003$ & 0.004 & $-0.173$ & 0.195 & $-0.012$ & $-0.021$ \\
2   & $-0.006$ & 0.007 & $-0.168$ & 0.189 & $-0.008$ & $-0.020$ \\
\hline\hline
\end{tabular}}
\end{table}

Previous analyses of beauty lifetimes~\cite{liferef,Bigis} have avoided
the proliferation of unknown hadronic parameters by making strong model
assumptions. The meson matrix elements of four-quark operators have
been estimated using the vacuum saturation approximation at a low
hadronic scale $\mu_{\rm had}$ which, according to (\ref{evol}), yields
$B_i\approx 1.0$ and $\varepsilon_i\approx -0.05$ at the scale $m_b$.
The baryon matrix elements have been estimated adopting the
valence-quark approximation $\widetilde B=1$ and assuming $r=O(1)$.
These assumptions lead to the model predictions
\begin{equation}
   \frac{\tau(B^-)}{\tau(B^0)} = 1 + (0.05\pm 0.03)\times
    \left( \frac{f_B}{200\,\mbox{MeV}} \right)^2 \,, \qquad
   \frac{\tau(\Lambda_b)}{\tau(B^0)} \approx 0.95 \,.
\label{nonsense}
\end{equation}
Whereas the first relation is in good agreement with the experimental
result in (\ref{taudata}), the predicted value for the ratio
$\tau(\Lambda_b)/\tau(B^0)$ is much larger than the experimental one.
This discrepancy is referred to as the ``$\Lambda_b$ lifetime puzzle''.

It has been advocated in Refs.~\cite{Bigis} that a precise measurement
of the lifetime ratio $\tau(B^-)/\tau(B^0)$ would allow for a
determination of the $B$-meson decay constant $f_B$ --- a proposal
which has been taken seriously by the DELPHI
Collaboration~\cite{DELfB}. Note that the ambiguity in the choice of
the matching scale alone implies a large uncertainty in the coefficient
in front of $f_B^2$ in (\ref{nonsense}), which is ignored in
Refs.~\cite{Bigis}. In addition, the numerical results for the
coefficients $k_i$ in Table~\ref{tab:1} show that assuming exact vacuum
saturation is not justifiable in the case of the ratio
$\tau(B^-)/\tau(B^0)$. To neglect in (\ref{master}) the deviation
$\Delta\varepsilon_1$ of the unknown parameter $\varepsilon_1$ from the
value $-0.05$ predicted by factorization at a low scale would only be
justified if $|\Delta\varepsilon_1|\ll 0.1$. The large-$N_c$ counting
rules of QCD do not allow us to impose such a strong restriction.

\begin{figure}[ht]
\epsfxsize=12cm
\centerline{\epsffile{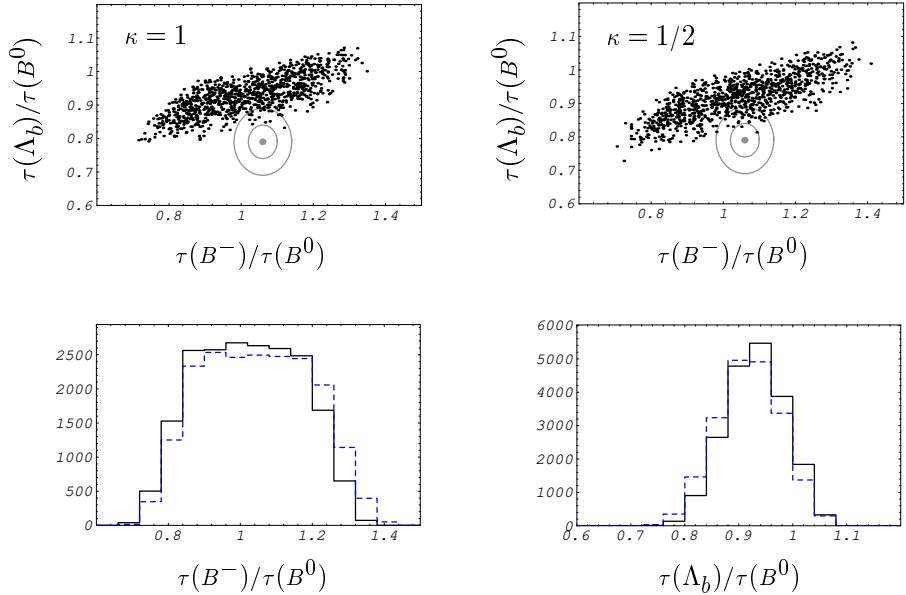}}
\caption{\label{fig:scatter}
Theoretical results for the lifetime ratios for random choices of the
hadronic parameters within the limits specified in the text. The
ellipses in the upper plots show the $1\sigma$ and $2\sigma$ contours
around the central experimental values. In the lower plots, the
projections on the individual lifetime ratios are shown. Full lines
correspond to $\kappa=1$, dashed ones to $\kappa=1/2$.}
\end{figure}

Our goal here is to investigate in a model-independent way how the
``predictions'' in (\ref{nonsense}) are altered if model assumptions
are avoided. To this end, we allow all hadronic parameters to vary
within reasonable limits and use the large-$N_c$ counting rules and
other theoretical prejudices only to restrict the allowed parameter
ranges. To be specific, we take random distributions inside the
following ranges: $B_i,\,\widetilde B\in [2/3,4/3]$, $\varepsilon_i\in
[-1/3,1/3]$, $r\in [0.25,2.5]$, and $\xi\in[0.8,1.2]$. For each
parameter set, the lifetime ratios $\tau(B^-)/\tau(B^0)$ and
$\tau(\Lambda_b)/\tau(B^0)$ are computed for the two perturbative
schemes with $\kappa=1/2$ and 1. The results are shown in
Fig.~\ref{fig:scatter}. In general, the scheme with $\kappa=1/2$ gives
results closer to the experimental data. There are two important
observations: first, only a small tail of the theoretical distribution
for the ratio $\tau(\Lambda_b)/\tau(B^0)$ reaches into the region
preferred by the experimental data; secondly, the predictions for the
ratio $\tau(B^-)/\tau(B^0)$ have an almost flat distribution between
0.85 and 1.25, indicating that without a better knowledge of the
hadronic parameters it is not possible to predict this ratio with good
precision. At present, not even the sign of the deviation from unity
can be predicted without model assumptions, in stark contrast with the
model-dependent result in (\ref{nonsense}).

\begin{figure}[p]
\epsfxsize=14cm
\centerline{\epsffile{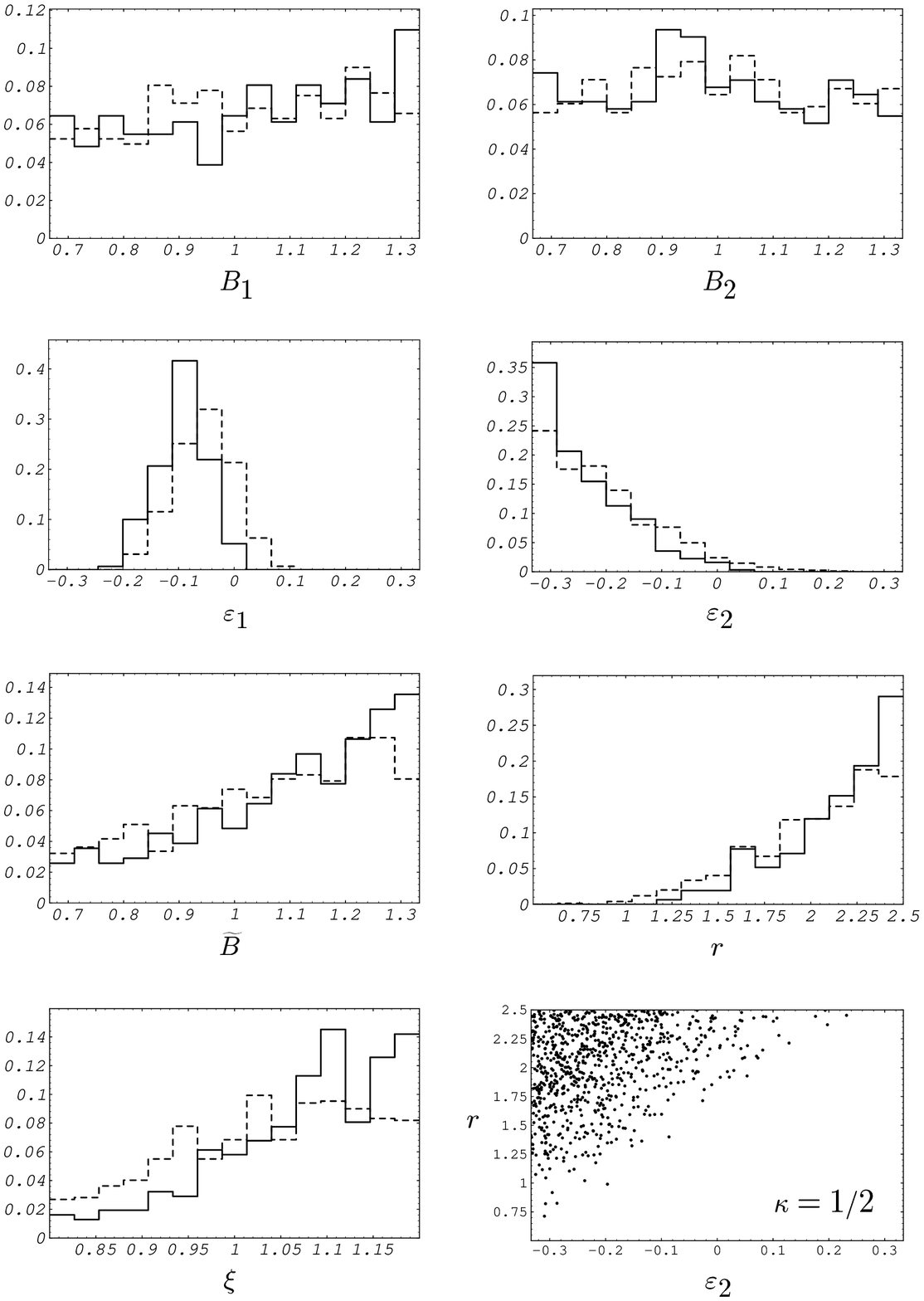}}
\caption{\label{fig:proj}
Distributions of the hadronic parameters for the simulations giving
results within $2\sigma$ of the central experimental values.}
\end{figure}

In the next step, we focus on the parameter sets which give results
within the $2\sigma$ ellipse around the central experimental values.
For these, the projected distributions for all hadronic parameters are
shown in Fig.~\ref{fig:proj}. The distributions of $B_1$ and $B_2$ and,
to a lesser extend, $\widetilde B$ and $\xi$ are rather flat,
indicating that these parameters are little restricted by the data. The
remaining three parameters are highly constrained, however. The
parameter $\varepsilon_1$, in particular, is determined by the
experimental data within a rather limited range: $\varepsilon_1\approx
-(0.1\pm 0.1)$, in good agreement with the expectation
$\varepsilon_1\approx -0.05$ based on the vacuum saturation
approximation. The distributions of the parameters $\varepsilon_2$ and
$r$ are strongly correlated. To reproduce the experimental data
requires to have $\varepsilon_2$ negative and $r\gsim 1$. The
correlation between the two parameters in shown in the last plot (for
the scheme with $\kappa=1/2$). If we assume that
$|\varepsilon_2|<0.15$, for instance, then $r$ must be larger than
about 1.25, which is higher than most expectations~\cite{Gube,Cola}. To
summarize, although some large hadronic parameters are needed to
explain the observed pattern of beauty lifetimes, a conventional
explanation of the $\Lambda_b$ lifetime puzzle cannot be excluded
before reliable field-theoretic calculations of the matrix elements of
four-quark operators become available.

\section{Conclusions and Implications}

The heavy-quark expansion, supplemented by the assumption of
quark-hadron duality, provides the theoretical framework for a
systematic calculation of the lifetimes and inclusive decay rates of
hadrons containing a heavy quark. Whereas this formalism works well for
the description of inclusive semileptonic decays, two potential
problems related to nonleptonic decays have become apparent in recent
years: the low experimental value of the lifetime ratio
$\tau(\Lambda_b)/\tau(B^0)$, and the low values of the semileptonic
branching ratio and charm yield in $B$ decays.

In order to obtain a detailed understanding of beauty lifetimes it is
necessary to go to order $1/m_b^3$ in the heavy-quark expansion, at
which the matrix elements of four-quark operators appear. They describe
the physics of nonspectator effects, i.e.\ contributions in which a
light quark in a beauty hadron is actively involved in the weak
interaction. We have presented a model-independent study of such
contributions introducing a minimal set of hadronic parameters, which
eventually may be determined using some field-theoretic approach such
as lattice gauge theory. We find that in $B$-meson decays the
coefficients of the colour-octet nonfactorizable operators are much
larger than those of the colour-singlet factorizable operators, and
therefore the contributions from the nonfactorizable operators cannot
be neglected. We also find that the theoretical predictions for
lifetime ratios still suffer from large perturbative uncertainties,
which could only be reduced if the short-distance coefficients in the
heavy-quark expansion were calculated to next-to-leading order. The
theoretical prediction for the ratio $\tau(B^-)/\tau(B^0)$ is in
agreement with experiment; however, our present ignorance about the
matrix elements of four-quark operators does not allow us to calculate
this ratio with an accuracy of better than about 20\%. As a
consequence, it is not possible to extract the $B$-meson decay constant
$f_B$ from a measurement of $\tau(B^-)/\tau(B^0)$.

The short $\Lambda_b$ lifetime remains a potential problem for the
heavy-quark theory. If the current experimental value persists, either
some hadronic matrix elements in the baryon and/or meson sector must be
larger than naive expectations, or local quark-hadron duality fails in
nonleptonic inclusive decays. In the latter case, the explanation of
the puzzle of the $\Lambda_b$ lifetime lies beyond our present
capabilities. We stress, however, that in view of our results it is
premature to conclude a failure of the heavy-quark expansion caused by
sizeable duality violations. There is a small region of parameter space
where the experimental data can be accommodated by theory.

If indeed the observed pattern of beauty lifetimes is caused by a
conspiracy of some large hadronic matrix elements, which are the
implications and tests of this scenario? A crucial test will be to
prove (or disprove) with a field-theoretical calculation that at least
one of the two parameters $\varepsilon_2$ and $r$ has the large value
required to explain the data. At present, efforts to compute these
parameters using lattice gauge theory are under way~\cite{priv}. An
important implication concerns the theoretical understanding of the
semileptonic branching ratio and charm yield in $B$ decays. These
issues are discussed at length elsewhere in these
Proceedings~\cite{Isi}. All theoretical calculations of these
quantities so far have neglected nonspectator
contributions~\cite{Alta}--\cite{Baga}. The most recent theoretical
predictions are~\cite{MNChris}
\begin{eqnarray}
   \hbox{B}_{\rm SL} &=& \cases{
    12.0\pm 1.0 \% ;& $\mu=m_b$, \cr
    10.9\pm 1.0 \% ;& $\mu=m_b/2$, \cr} \nonumber\\
   n_c &=& \cases{
    1.20\mp 0.06 ;& $\mu=m_b$, \cr
    1.21\mp 0.06 ;& $\mu=m_b/2$, \cr}
\label{Bslnc}
\end{eqnarray}
with correlated errors. Notice that the semileptonic branching ratio
has a stronger scale dependence than $n_c$. By choosing a low
renormalization scale, values $\hbox{B}_{\rm SL}<11\%$ can be
accommodated. Currently, the experimental values of $B_{\rm SL}$ and
$n_c$ are still controversial; however, the value of $n_c$ tends to lie
below the theoretical predictions.

Nonspectator contributions can change the predictions for the
semileptonic branching ratio significantly; however, they have a
negligible effect on $n_c$. Irrespective of the precise values of the
hadronic parameters, there is a model-independent relation $\Delta
n_c\approx\Delta B_{\rm SL}\approx 0$--1\%~\cite{MNChris}, i.e.\ the
relative effect on $B_{\rm SL}$ is ten times larger than that on $n_c$.
A large negative value of $\varepsilon_2$ implies a positive
contribution to $B_{\rm SL}$, which is not preferred by the data. This
restriction favours the solution where the $\Lambda_b$ lifetime puzzle
is solved by a large value of the baryon parameter $r$.

The fact that the charm yield $n_c$ is very little affected by
nonspectator effects is interesting. To reduce $n_c$ significantly
below the values quoted in (\ref{Bslnc}) requires an anomalously large
branching ratio for charmless nonleptonic $B$ decays. At present, a
value of $\hbox{B}(B\to X_s+\mbox{no charm})$ several times larger than
predicted by the Standard Model is not excluded by the data. Such
interesting scenarios are explored elsewhere in these
Proceedings~\cite{Alex}.

\section*{Acknowledgements}

Much of the work presented here has been done in a most enjoyable
collaboration with Chris Sachrajda, which is gratefully acknowledged. I
would also like to thank Tom Browder, Sandip Pakvasa and their staff
for making this conference most pleasant and inspiring.

\end{document}